# Catalyst Diffusion Transformer: Generative Inverse Design of Heterogeneous Catalysts

Hayoung Doo[1, 2, 3, †], Dong Hyeon Mok[4, †], Seoin Back[3, 5, 6, 7 *], Jonggeol Na[1, 2, 3, *]

[1]Department of Chemical Engineering and Materials Science, Ewha Womans University, Seoul 03760, Republic of Korea

[2]Department of Chemical Engineering, Graduate Program in System Health Science and Engineering, Ewha Womans University, Seoul, 03760, Republic of Korea

[3]Institute for Multiscale Matter and Systems (IMMS), Ewha Womans University, Seoul 03760, Republic of Korea

[4]Department of Chemical and Biomolecular Engineering, Institute of Emergent Materials, Sogang University, Seoul 04107, Republic of Korea

[5]KU-KIST Graduate School of Converging Science and Technology, Korea University, Seoul 02841, Republic of Korea

[6]Department of Integrative Energy Engineering, Korea University, Seoul 02841, Republic of Korea

[7]Center for Hydrogen and Fuel Cells, Korea Institute of Science and Technology (KIST), Seoul 02792, Republic of Korea

**Correspondence and requests for materials should be addressed to**

S.B. (email: sback@korea.ac.kr), J.N. (email: jgna@ewha.ac.kr)

## ABSTRACT

The vast chemical design space and complex, interdependent design variables make catalyst discovery for targeted properties highly labor- and resource-intensive. Although generative models have emerged as a promising solution, existing approaches are generally limited to single-property conditioning or narrow chemical spaces. Here, we present Catalyst Diffusion Transformer (CatDiT), a unified framework for inverse catalyst design that generates valid and novel structures ranging from intermetallic alloys to oxide surfaces. By learning compressed latent representations, CatDiT enables efficient training and rapid sampling while supporting simultaneous conditioning on adsorbate type, binding energy, and catalyst class. The model provides reliable control of discrete properties and directional control of continuous properties, enriching candidate pools for reaction-specific catalyst discovery. As a representative application, multi-conditional generation for the nitrogen reduction reaction (NRR) yields 28 density functional theory (DFT)-relaxed alloy candidates that satisfy the target activity window and lie above the pure-metal *N–*H scaling line, corresponding to a ~1.5-fold enrichment over the source distribution. These results establish CatDiT as a practical and scalable approach for property-directed catalyst inverse design and targeted catalyst generation.

## Introduction

Functional properties lie at the core of materials science, and discovering materials that satisfy targeted performance requirements remains a central pursuit. Conventional computational approaches rely on screening libraries of known structures and ranking candidates based on desired properties. Recent advances in machine learning have significantly reduced the cost of this brute-force evaluation process, yet screening methods remain constrained by predefined candidate libraries and are unable to fully explore the immense chemical and structural search space of possible materials. Overcoming this limitation requires moving beyond fixed candidate sets and treating materials discovery as an inverse design problem, in which structures are generated to satisfy specific functional constraints. Generative modeling has therefore emerged as a promising strategy for inverse materials design, demonstrating success in molecules[1–3], crystals[4,5], proteins[6,7], and metal-organic frameworks[8,9].

Extending this paradigm to heterogeneous catalysts is particularly important yet challenging. Catalytic activity depends on the coupled effects of composition, surface facet, adsorbate identity, and coverage on extended periodic surfaces, creating a vast combinatorial design space involving both local adsorbate–surface interactions and long-range structural features. Traditionally, this space has been explored using density functional theory (DFT)-based adsorption energy descriptors, but exhaustive screening is impractical because of the exploding number of possible combinations[10]. While machine-learning interatomic potentials (MLIPs) and other surrogate models have significantly improved computational efficiency[11], they remain largely limited to evaluating existing structures rather than designing new catalyst candidates.

To tackle this challenge, generative approaches for catalyst design have emerged. Early efforts focused on individual subproblems such as adsorbate placement and stable surface search[12,13]. Later studies generated complete slab–adsorbate structures by predicting three-dimensional atomic coordinates of the joint system conditioned on adsorbate type[14–16]. More recently, property-guided inverse design has attracted increasing attention. MAGECS[17] optimized alloy surfaces for target $CO_2$ reduction adsorption energies through latent-space exploration, while Yang et al.[18] steered a descriptor-guided diffusion model with a target descriptor to generate bimetallic alloys for ammonia decomposition. Despite these advances, existing property-conditioned methods are largely limited to reaction-specific datasets and narrow compositional spaces. Across the field, catalyst inverse design still lacks a unified framework that combines simultaneous multi-property conditioning across a broad chemical space.

Diffusion-based models are well-suited to address this limitation owing to their expressive power and flexible conditioning capabilities[19] and have driven much of the recent progress described above. However, directly applying diffusion in atom-coordinate space becomes computationally prohibitive as system size grows. Slab–adsorbate configurations typically contain tens to hundreds of atoms, far exceeding the scale of molecules or bulk unit cells targeted by most existing generative models. Latent diffusion addresses this challenge by first encoding structures into a compact, chemically meaningful representation[20] and then performing the generative process entirely in this learned latent space, making generation tractable for large periodic systems without sacrificing structural expressiveness.

Motivated by these considerations, we propose CatDiT, a latent diffusion framework for the inverse design of heterogeneous catalysts. Our main contributions are threefold. First, CatDiT encodes catalyst structures into an SE(3)-equivariant latent space and performs flow matching to achieve state-of-the-art structure validity, uniqueness, and relaxation energy stability among existing slab-adsorbate generative models. Second, it unifies multiple property conditions within a single conditioning framework, enabling both discrete and continuous property-guided generation. Third, CatDiT expands the accessible chemical space from intermetallic surfaces to oxides and demonstrates the ability to generate structurally feasible, diverse, and novel catalyst candidates across all catalyst classes.

## Results

### CatDiT framework

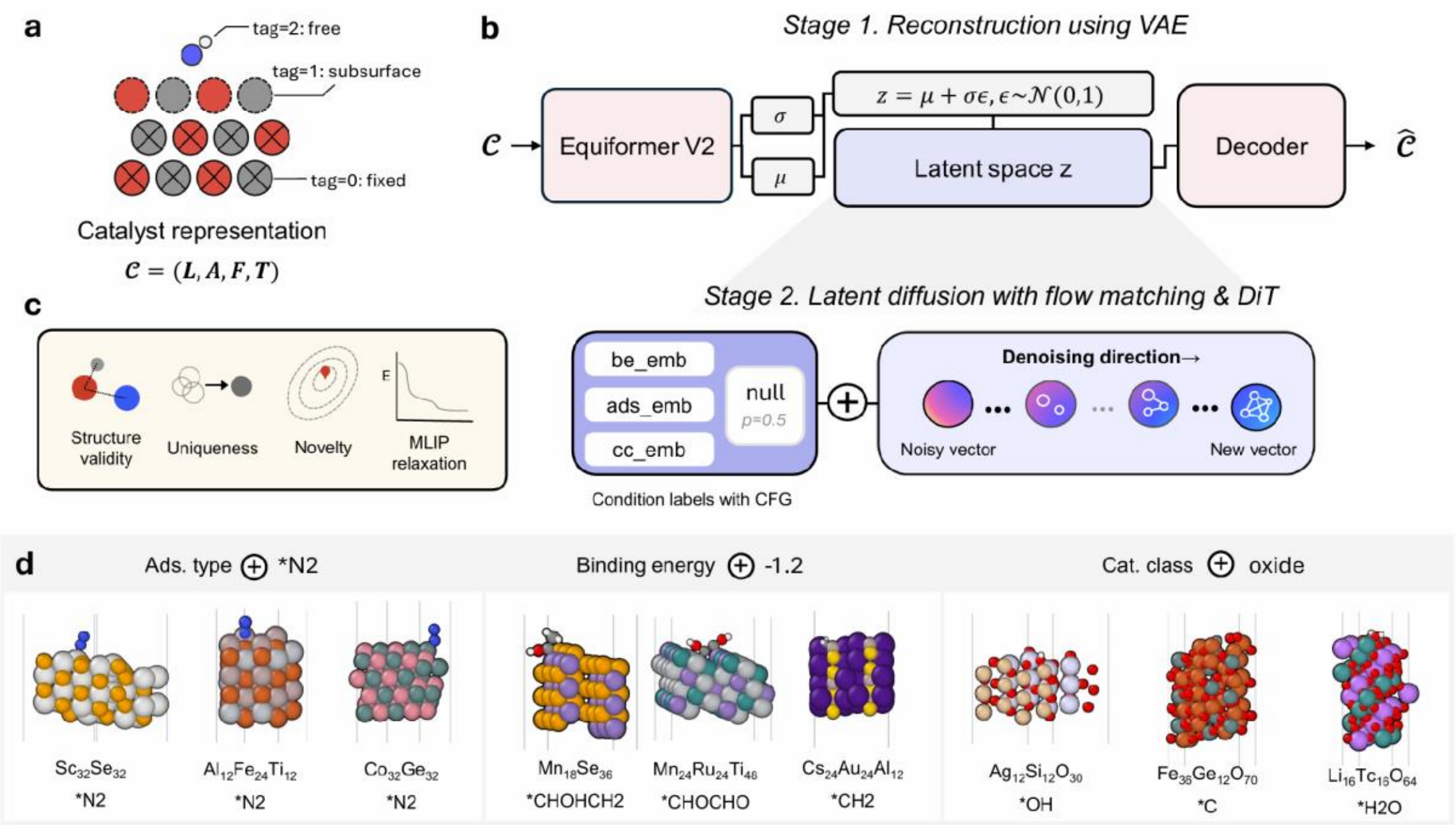

**Figure 1. Overview of CatDiT framework. a** Catalyst input represented as a graph, where the tuple ($\boldsymbol{C}$) comprises atom types, lattice information, fractional coordinates, and atomic tags indicating whether each atom is fixed or free. **b** Two-stage latent diffusion architecture for conditional catalyst generation, consisting of a VAE for SE(3)-equivariant latent representation learning with EquiformerV2[21] and DiT for flow matching in latent space. Three property labels enable conditional generation through CFG. **c** Primary metrics for evaluating generation quality. **d** Visualization of generated catalyst structures under specified conditions.

As conceptually illustrated in Figure 1, CatDiT is a two-stage latent diffusion model built on ADiT[22] and employing EquiformerV2[21] as the encoder to ensure equivariant representation. The model incorporates three property labels, namely adsorbate type, catalyst class, and binding energy, to enable both unconditional and conditional generation within a single model. In the first stage, a variational autoencoder (VAE) learns compact latent representations by reconstructing catalyst structures. The trained VAE is then frozen to provide latent embeddings for corruption and denoising, while the diffusion Transformer (DiT) performs latent generation through flow matching. Property labels are injected via classifier-free guidance (CFG) to steer generation toward desired characteristics. Finally, generated structures are evaluated for structure validity, uniqueness, and novelty, and relaxed using an MLIP to assess minimum stability.

All CatDiT variants share the same architecture and differ only in their property conditioning and training datasets (Supplementary Table 1). Implementation details are provided in the Methods section.

## Generation performance comparison with baselines

Evaluating generative models for catalyst structures poses challenges distinct from those for bulk crystals. Stability metrics developed for stoichiometric bulk phases are not directly applicable to cleaved, non-stoichiometric surfaces, and diversity assessment must distinguish adsorbate configurations from the underlying slab. Nevertheless, several minimum criteria remain valid across systems. Generated structures should be physically plausible, non-redundant, and sufficiently close to a local minimum to provide meaningful starting points for downstream relaxation. We therefore benchmark CatDiT against CatGPT[15] and CatFlow[16], two models for complete slab–adsorbate generation, using six metrics of generation quality and relaxation stability. Overall, CatDiT achieves the highest structure validity and uniqueness and the lowest relaxation energy change (Table 1), demonstrating the effectiveness of latent diffusion for generating physically plausible slab–adsorbate structures.

The high structure validity and uniqueness are likely attributable to the geometry-aware latent representations learned by the SE(3)-equivariant encoder, which reduce sensitivity to arbitrary coordinate choices and discourage redundant generation of geometrically similar structures[23,24]. Although CatDiT requires more relaxation steps on average than the baseline models, its comparatively lower $\Delta E_{sys}$ indicates smaller energetic corrections during relaxation and greater consistency between initial and relaxed structures. These results should be interpreted as framework-level benchmarks rather than strictly controlled ablation studies, as the baseline models differ in training data scale, task formulation, and preprocessing protocols. Notably, our preprocessing removes dissociated and desorbed adsorbate configurations, restricting training to physically meaningful bound states and potentially improving structure–energy consistency.

**Table 1. Comparison with baseline models.** Metrics and results for CatGPT and CatFlow are taken from the CatFlow study. CatDiT is evaluated using the same evaluation code and the UMA[25]-based relaxation protocol.

| Method | Structure validity (%) | Uniqueness (%) | Comp. diversity | $\Delta E_{sys}$ (eV) | Conv. steps | Conv. rate (%) |
|---|---|---|---|---|---|---|
| CatGPT | 92.67 | 79.91 | 15.0655 | 33.1267 | 121.0629 | 98.26 |
| CatFlow | 97.33 | 94.69 | 15.0724 | 28.0060 | **115.0534** | **99.22** |
| CatDiT | **98.32** | **99.26** | **15.0805** | **23.0163** | 131.1208 | 99.12 |

## Novelty, diversity, and distribution coverage

Beyond the metrics in Table 1, coverage and property distributions are important for assessing whether a generative model reproduces plausible structures spanning the ground-truth (GT) distribution. We evaluate structural and compositional coverage using CrystalNN and Magpie fingerprints[4,25], respectively, and quantify property distributions with the Earth mover's distance (EMD)[26] (Methods section). As shown in Figure 2a, the distributions of generated structures closely match the validation set in lattice volume, density, adsorption energy, and number of elements for OC20[27], whereas OC22[28] exhibits slightly unstable coverage due to its smaller sample size.

Notably, high coverage alone does not ensure meaningful generation. A model that merely memorizes and regurgitates the GT distribution would achieve perfect coverage while offering no new candidates. The structure panel of Figure 2a suggests that CatDiT avoids this issue, exhibiting slightly larger nearest-neighbor distances to the validation set than the training reference. This indicates that the model introduces structural variations rather than replicating known configurations.

For bulk materials, `StructureMatcher`[29] is commonly used to assess novelty by identifying structures present in the training set. However, its application to catalyst systems is challenging. First, "identical catalysts" are difficult to define because catalyst surfaces cleaved from the same bulk composition along different crystallographic planes can exhibit substantially different catalytic properties. Second, catalyst structures containing tens to hundreds of atoms are highly sensitive to minor variations, with slight deviations in atom count or atomic coordinates often leading to classification as distinct structures. Combined with the difficulty of selecting appropriate tolerance parameters, `StructureMatcher` tends to overestimate novelty in catalyst systems. To address this limitation, we adopt continuous distance measures for novelty evaluation[30]. Structural similarity is quantified using $d_{AMD}$, the $L_\infty$ distance between average minimum distance (AMD) vectors[31], which satisfies both invariance under isometries and Lipschitz continuity—two properties that `StructureMatcher` lacks. Even structures sharing the same bulk composition and similar facets exhibit $d_{AMD}$ differences on the order of $10^{-2}$ (Supplementary Table 2, Supplementary Figure 1), demonstrating that the metric resolves structural variations at the catalyst-surface level. Compositional similarity is measured using $d_{magpie}$, the L2 distance between Magpie fingerprints.

Simply exploring outside the training distribution does not necessarily imply meaningful discovery, as truly novel compositions must also be physically realizable. We therefore cross-reference each novel structure against entries in the Materials Project[32]. Since catalyst slabs cleaved along different crystallographic planes may deviate from the ideal stoichiometries of their parent bulk phases owing to slab thickness, termination, and surface reconstruction, we employ near-composition matching rather than exact one-to-one matching. Specifically, each slab is matched to Materials Project bulk entries with the same elemental constituents or similar stoichiometry. For example, $Cu_{24}Pd_8$ is reduced to $Cu_3Pd$, whereas off-stoichiometric $Cu_{31}Pd_{10}$ is considered compositionally close to $Cu_3Pd$, as described in the Supplementary Methods. We then determine whether at least one matched bulk entry has $E_{hull}$ < 0.1 eV/atom.

The presence of a near-stable bulk entry with a similar composition indicates that the generated slab lies within a low-energy bulk composition space. However, it neither guarantees that the surface can be cleaved from the corresponding bulk phase nor confirms its thermodynamic stability or experimental synthesizability. Conversely, the absence of a near-stable Materials Project match does not invalidate the generated slab but leaves its bulk recoverability and stability unresolved. Figure 2b summarizes the continuous novelty analysis. Of the 9,832 valid generated structures, 35.9% share compositions with the training set, although this does not imply structural redundancy. As shown in the scatter plot, structures with compositions identical to those in the training set ($d_{magpie}$ = 0) still exhibit nonzero $d_{AMD}$ values, confirming that CatDiT generates new surface configurations rather than reproducing existing structures. Among all generated structures, 10.2% have novel compositions with at least one near-stable bulk match in the Materials Project, while 53.3% remain unresolved under this criterion. Figure 2c exhibits the nine representative novel structures that highlight the chemical diversity of the generated space, spanning transition metals, main-group elements, oxides, and various bound adsorbates. Notably, chemical systems, such as Ti–Si–Tc and Se–Cd, are absent from the training set but appear in the OC20 OOD-test subset, suggesting the model's extrapolation capability.

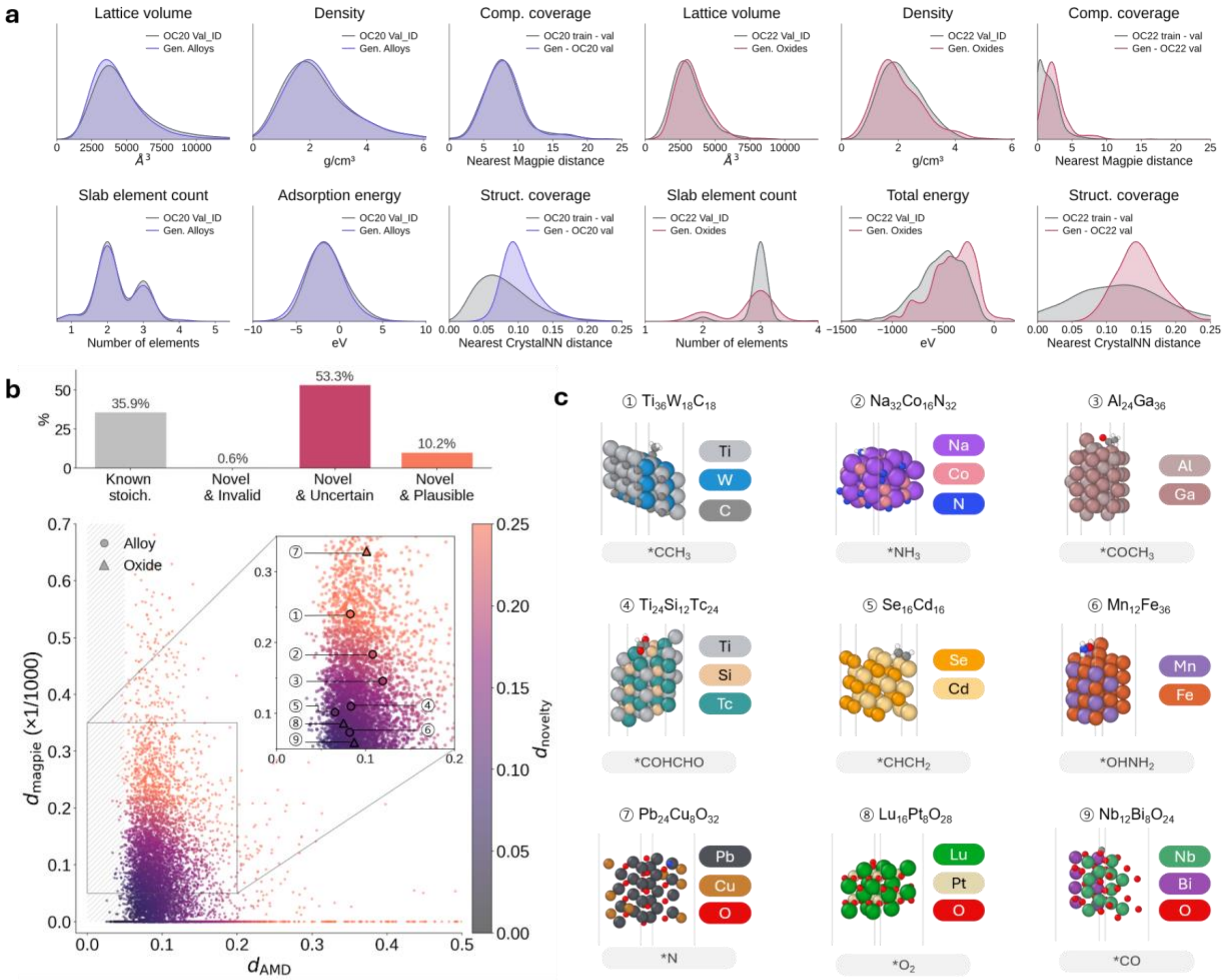

**Figure 2. Exploration of the chemical space covered by CatDiT-generated structures. a** Property distributions of generated structures (purple and red) compared with the Val-ID (gray) sets of OC20 and OC22. Coverage is evaluated using nearest-neighbor fingerprint distances to Val-ID, with Train → Val (gray) distances shown as a reference. **b** Top: novelty classification of 9,832 structurally valid structures based on near-composition matching. "Known" represents structures compositionally matched to the training set; novel structures are classified as "Plausible" if a near-stable Materials Project entry with $\boldsymbol{E_{hull}}$< 0.1 eV/atom

exists, “Uncertain” if no such match is found or $\boldsymbol{E_{hull}} \geq 0.1$ eV/atom, and “Invalid” for non-catalytic compositions (e.g., unary non-metal, noble gas, or halogen-only compounds). Bottom: continuous novelty analysis based on structural ($\boldsymbol{d_{AMD}}$) and compositional ($\boldsymbol{d_{magpie}}$) distances to the nearest training structure. Color indicates the combined Euclidean novelty score. The inset enlarges the dense region, and numbered points correspond to the structures shown in c. **c** Nine representative novel structures spanning diverse chemistries and adsorbates. Structures 4–6 are classified as Novel & Plausible, whereas the remainder are Novel & Uncertain.

**Conditional generation on discrete variables**

As mentioned earlier, unconditional generation samples from the overall training distribution, requiring post hoc filtering to obtain structures with specific adsorbate or catalyst composition. Conditional generation avoids this overhead by directly steering the model toward desired properties in a single pass. Building on the base CatDiT model, we train CatDiT-A and CatDiT-C (Supplementary Table 1) for the conditional generation of two discrete variables—adsorbate identity and catalyst class, respectively. To evaluate whether each model faithfully generates structures that match the specified conditions, we assess condition fidelity, defined as the fraction of generated structures satisfying the corresponding condition.

**Table 2. Conditional generation results of CatDiT-C across target catalyst classes.** Generated structures are evaluated across five catalyst classes, ranging from intermetallic alloys to oxide surfaces. Catalyst classes are defined by the elemental composition of the slab. Condition fidelity denotes the fraction of generated samples that satisfy the target catalyst class.

| Class | Criteria | Condition fidelity | Structure validity | Convergence rate |
|---|---|---|---|---|
| 0: Intermetallic / Unary | ≥1 metal in the slab | 100.00% | 98.00% | 98.37% |
| 1: Metalloid | Metalloid in the slab | 98.77% | 97.60% | 98.77% |
| 2: Non-metal | Non-metal anion in the slab | 99.58% | 95.80% | 96.87% |
| 3: Halides | Halogen in the slab | 100.00% | 97.20% | 91.15% |
| 4: Oxide | O in the slab | 100.00% | 92.40% | 92.86% |

We first evaluate class-conditioned generation across five catalyst classes: intermetallic/unary, metalloid, non-metal, halide, and oxide systems. As shown in Table 2, condition fidelity exceeds 98.7% for all classes, demonstrating that a single class label is sufficient to guide generation toward the intended catalyst class. This capability extends controllable generation from intermetallic surfaces to oxides, covering a broader chemical space than previous alloy-centric catalyst generators. Structure validity remains high but varies more across classes, from 98.0% for intermetallic/unary surfaces down to 92.4% for oxides. The lower validity for oxides likely reflects the greater structural complexity of oxygen-containing surfaces in OC22, which are more prone to surface reconstruction and adsorbate dissociation. Condition fidelity remains high even for these structurally demanding classes, indicating that class-conditioned generation extends to chemistries underrepresented in prior generative catalyst design studies.

Turning to a higher-dimensional discrete condition, we evaluate CatDiT-A on adsorbate-conditioned generation across 64 species retained from the 82 species in OC20 after excluding labels absent from the training set. Figure 3a shows strong diagonal dominance in the confusion matrix, indicating high condition fidelity. Compared with the catalyst class, adsorbate identity is a more fine-grained condition that requires control over both elemental composition and molecular connectivity, leading to greater variation in fidelity across species. It is worth noting that CatGPT and CatFlow achieve near-perfect match rates by treating adsorbates as direct inputs, whereas CatDiT injects adsorbate identity as a property label through CFG. This softer conditioning mechanism does not enforce exact matches but provides explicit control over the fidelity–diversity tradeoff through the guidance scale, $w$.

To investigate the sources of fidelity variation, we conduct a correlation analysis and identify three dominant factors (Figure 3b). Atom count is the strongest predictor: monoatomic adsorbates, such as *O and *N, achieve fidelities above 70%, whereas species with eight or more atoms fall below 42%, reflecting the greater structural degrees of freedom of larger molecules. Training frequency also correlates positively with fidelity, although larger adsorbates are simultaneously underrepresented in the training data, compounding the difficulty. Isomer count is another important factor, as species with identical molecular formulas but different connectivity exhibit higher inter-isomer confusion. Indeed, the most frequent misassignments are constitutional isomers of the target adsorbate (Supplementary Figure 2 for the full per-species breakdown). The strong correlation between isomer count and confusion rate suggests that distinguishing isomers requires a topological understanding of molecular connectivity that is not fully captured by the current conditioning mechanism.

Crucially, this fidelity limitation does not propagate to structural quality. CatDiT-A maintains a mean structural validity of 92.8% across all conditioned generations (Figure 3c), indicating that errors in adsorbate identity do not compromise the validity of the generated surfaces. Fidelity and validity therefore reflect distinct aspects of model performance, suggesting that further improvements should target the conditioning mechanism rather than the underlying generation strategy.

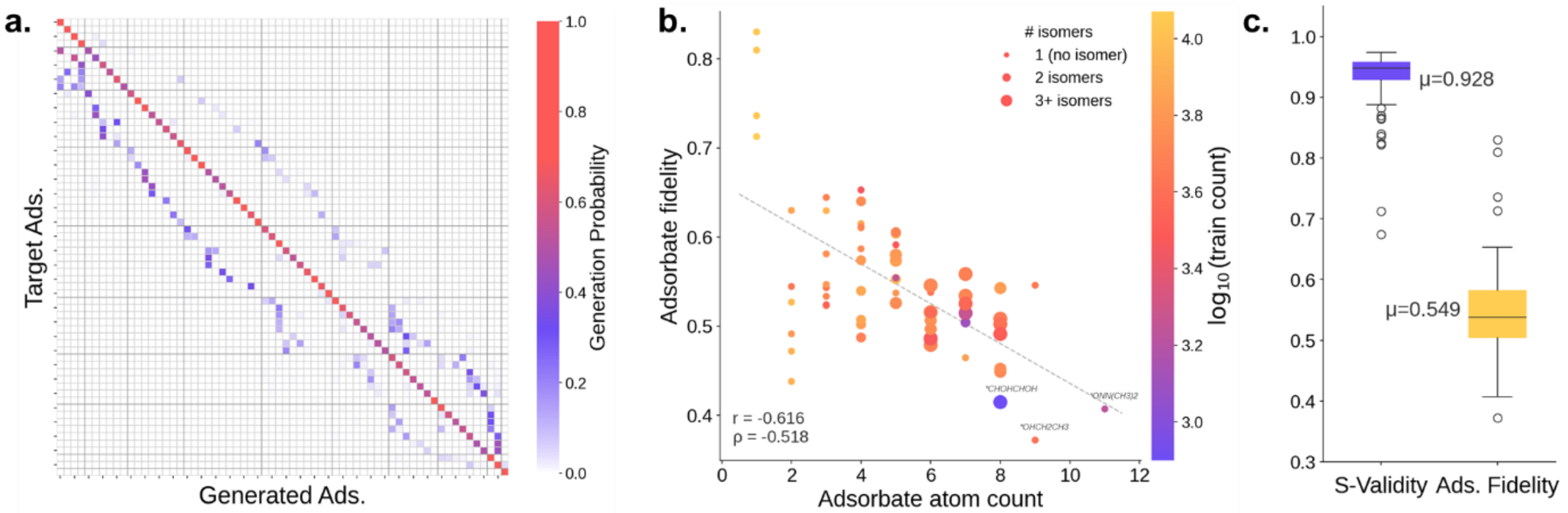

**Figure 3. Adsorbate-conditioned generation performance of CatDiT-A. a** Confusion matrix of condition fidelity across 64 adsorbate species. Rows and columns correspond to target and generated adsorbates, respectively; diagonal entries (red) indicate correct generation and off-diagonal entries (purple) indicate misassignments. **b** Correlation of adsorbate fidelity with atom count, training frequency, and isomer count. The dashed line denotes the linear fit (Pearson r = –0.616; Spearman ρ = –0.518). **c** Distributions of structure validity and adsorbate fidelity across all conditioned generations.

**Conditional generation on continuous variables**

Given the central role of adsorption energy as a descriptor for screening active surfaces and identifying intermediate binding regimes, we next evaluate whether CatDiT-B can condition generation on this continuous variable. Unlike discrete labels, continuous conditioning poses a one-to-many challenge, as a single $E_{ads}$ value may correspond to diverse structural configurations. The model must therefore sample from a broad structural manifold while steering generation toward the desired energy regime, making this task fundamentally more difficult than matching a categorical label. CatDiT-B is designed for this setting by incorporating adsorption energy, also referred to here as binding energy, as a conditioning signal.

As shown in Figure 4, the MLIP-predicted $E_{ads}$ distribution shifts monotonically with the target values, indicating directional control by the conditioning signal. However, the generated distributions remain broad, and conditioning accuracy varies across the energy range. Energy control is strongest in the data-rich region between −4 and +2 eV, where up to 36% of samples fall within a ±0.5 eV tolerance. In contrast, both target alignment and tolerance fractions decrease at the sparsely sampled distribution tails (±6, ±8 eV). This behavior indicates that continuous conditioning is most reliable within the learned data manifold and becomes less precise in underrepresented energy regimes. This does not suggest that the model generates arbitrary structures at extreme targets; rather, it indicates that conditioning precision degrades where training data are sparse.

Importantly, conditioning on extreme energy values does not substantially compromise structural quality. Structure validity remains above 93% across nearly the entire target range, with similar MLIP convergence behavior (Figure 4b). Although absolute energy precision is limited, with the ±0.2 eV tolerance fraction peaking at ~15%, the model consistently provides directional control over the generated $E_{ads}$ distributions. Given the one-to-many nature of adsorption energy, this level of control offers a useful prior for an underdetermined inverse design problem and represents a meaningful early step toward property-guided catalyst generation. In practice, it can serve as a prefilter to concentrate candidates within a target binding energy range before downstream MLIP or DFT relaxation, thereby reducing screening costs.

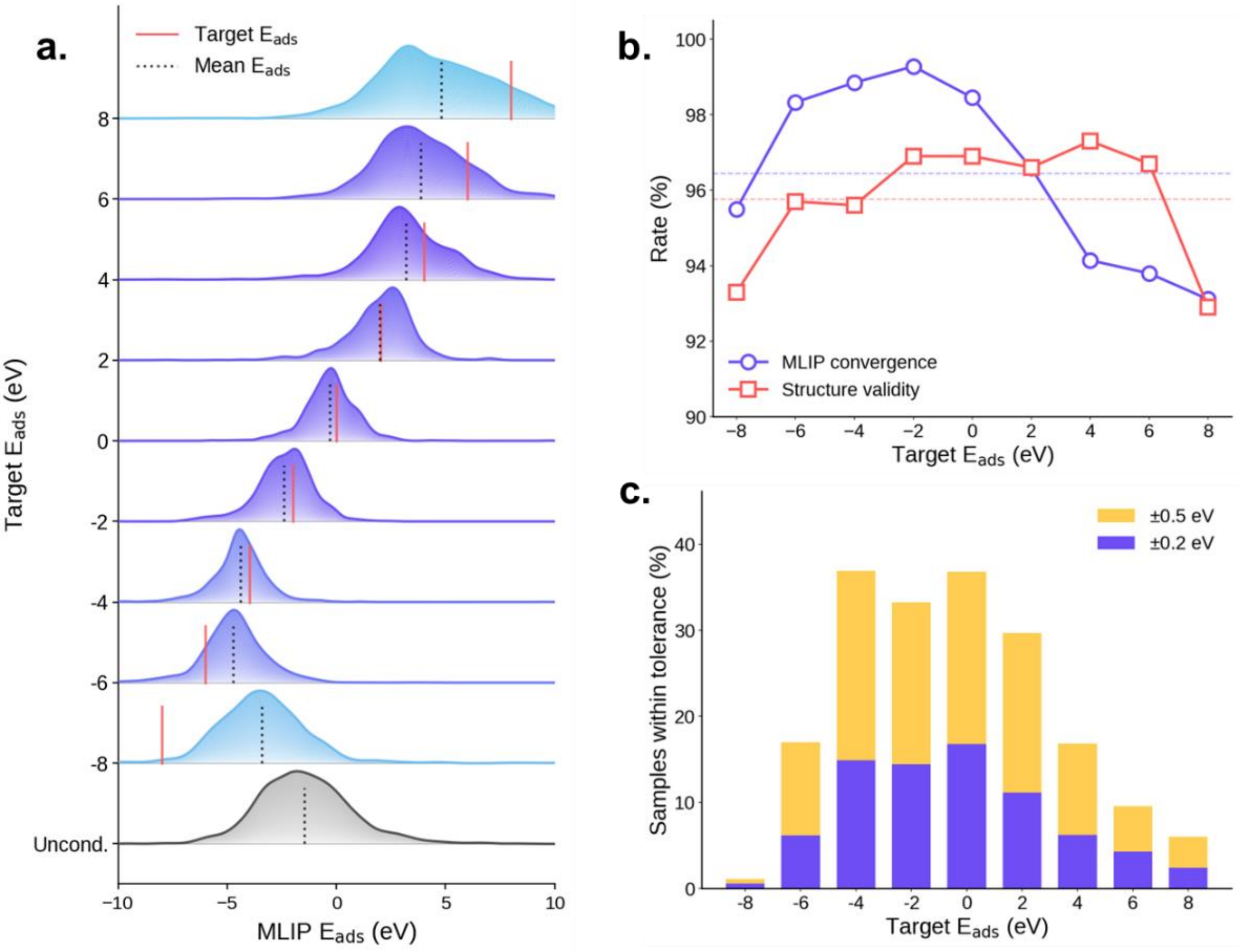

**Figure 4. Conditional generation performance of CatDiT-B across target adsorption energies. a** Ridge plot of MLIP-predicted $E_{ads}$ distributions for each target value, with the unconditional distribution shown in gray. Color indicates structure validity at each target, ranging from light blue (low) to dark purple (high). **b** Structure validity and MLIP convergence rate as a function of target $E_{ads}$; dashed lines denote the mean values across all targets. **c** Fraction of generated structures within ±0.2 eV (purple) and ±0.5 eV (yellow) of the target values.

**Multi-conditional generation for NRR applications**

To this end, the key capability for reaction-specific catalyst discovery is the simultaneous conditioning of catalyst structures on both adsorbate identity and binding energy. This enables the direct generation of candidates that stabilize intermediates with the desired adsorption strength. This substantially improves search efficiency by concentrating generated candidates within the reaction-relevant regime from the outset, reducing the need for exhaustive post hoc filtering.

As a representative application, we consider the nitrogen reduction reaction (NRR), an electrochemical route for ammonia synthesis that offers a potential alternative to the Haber-Bosch process. A major challenge in conventional NRR is the activity−selectivity tradeoff. On transition-metal surfaces, the binding energies of *N and *H are often linearly correlated, causing catalysts that stabilize nitrogen-containing intermediates to also bind hydrogen strongly[33–35]. Although pure-metal catalysts have been widely used to establish NRR descriptors and scaling relations, many thermodynamically favor HER over NRR because their HER limiting potentials are typically closer to zero than their NRR limiting potentials[35]. We therefore employ multi-conditional generation to identify alloy catalysts capable of breaking the scaling behavior of pure metals.

Key NRR intermediates, including $*N_2H$, *NH, and $*NH_2$, bind to catalyst surfaces through nitrogen, allowing their energetics to be approximated from $\Delta E_{N^*}$ using established scaling relations[25]. Although these relations may be weakened or modified in alloy systems, we assume that they remain useful for screening[36]. We therefore adopt $\Delta E_{N^*}$ as an activity descriptor and $\Delta E_{H^*}$ as a selectivity descriptor against HER competition. Following Skúlason et al.[33] and assuming a dissociative pathway on flat surfaces, we set the optimal activity target to –1.0 eV, corresponding to the approximate peak of the NRR volcano plot. The exact position of the volcano peak depends on the reaction mechanism (dissociative vs. associative), surface type (flat vs. stepped), and literature source, and is typically reported between –0.5 and –1.0 eV.

Using CatDiT-AB, we generated 10,000 samples conditioned on adsorbate *N and a binding energy of –1.0 eV. Of these, 9,704 were structurally valid, and 7,919 satisfied the target adsorbate condition. After MLIP relaxation for both *H and *N adsorption, 7,621 structures converged successfully for both ads-slab and clean slab systems. Among them, 721 candidates simultaneously satisfied the ±0.2 eV target $E_{ads}$ tolerance and lay above the pure-metal scaling line. This criterion serves as a proxy for improved NRR selectivity over HER, as candidates in this region are expected to weaken H binding relative to N binding compared with pure-metal surfaces[33].

Prioritizing DFT validation under computational cost constraints, we further filtered candidates by selecting structures containing literature-supported NRR-relevant elements and no more than 60 atoms. To broaden the search beyond conventional transition-metal motifs, we defined a set of 21 elements appearing in experimentally reported or computationally proposed NRR-relevant materials. These include i) transition-metal elements previously investigated for NRR (Mo, Fe, Ru, V, W, Re, Nb, Ti, Os, and Ta)[33,37], ii) emerging main-group p-block elements (Bi, Sb, Sn, B, Ga, In, and Pb)[38,39], iii) rare-earth elements which have been incorporated into intermetallic ammonia-synthesis catalysts (Y and Sc)[40,41], and iv) Mars–van Krevelen nitride-active elements (Zr and Hf)[42]. One caveat is that element membership was used solely as a prioritization criterion rather than as evidence of NRR activity. Applying these criteria yielded 167 candidates for DFT validation.

Among the 167 candidates selected for DFT validation, 81 converged for the slab, *N, and *H configurations within 12 hours (Figure 5). The remaining structures were considered incompletely optimized because they did not satisfy the force-convergence criterion within the allotted wall time and were not counted as hits, making the reported hit rate a conservative lower bound. Of the converged structures, 29 satisfied the target activity window, and 28 additionally met the pure-metal scaling-line criterion, $\Delta E_H > 0.22\Delta E_N - 0.36$. These results demonstrate that multi-conditional generation can produce reaction-specific catalyst candidates that simultaneously satisfy activity and selectivity requirements. When normalized by the total number of generated structures, this corresponds to 29 DFT-validated hits per 10,000 samples. In comparison, structures within the target window in the OC20 source distribution, defined as *N adsorbates with $E_{ads}$ between –1.2 and –0.8 eV, occur at a frequency of ~0.2%, or 20 per 10,000 structures. This corresponds to a ~1.5-fold enrichment. Notably, this lower-bound enrichment was achieved despite validating only a subset of candidates under a strict convergence criterion.

Interestingly, all 28 candidates satisfying both target criteria were confirmed to be structurally distinct from the training data. Within this subset, several candidates exhibit motifs consistent with established NRR chemistry, including Co–Mo and Fe–Mo combinations. The validated set also contains non-traditional slab-level motifs, such as Au–Y, Y–Zn, Fe–Zr–Sb, and Ga–Hf–Pt, which incorporate literature-supported NRR-relevant elements in chemically plausible compositions. These motifs should not be interpreted as stable bulk phases or readily synthesizable catalysts. Instead, they represent computational leads for further validation, including assessments of bulk recoverability, phase stability, adsorption-site sampling, and complete reaction-pathway calculations.

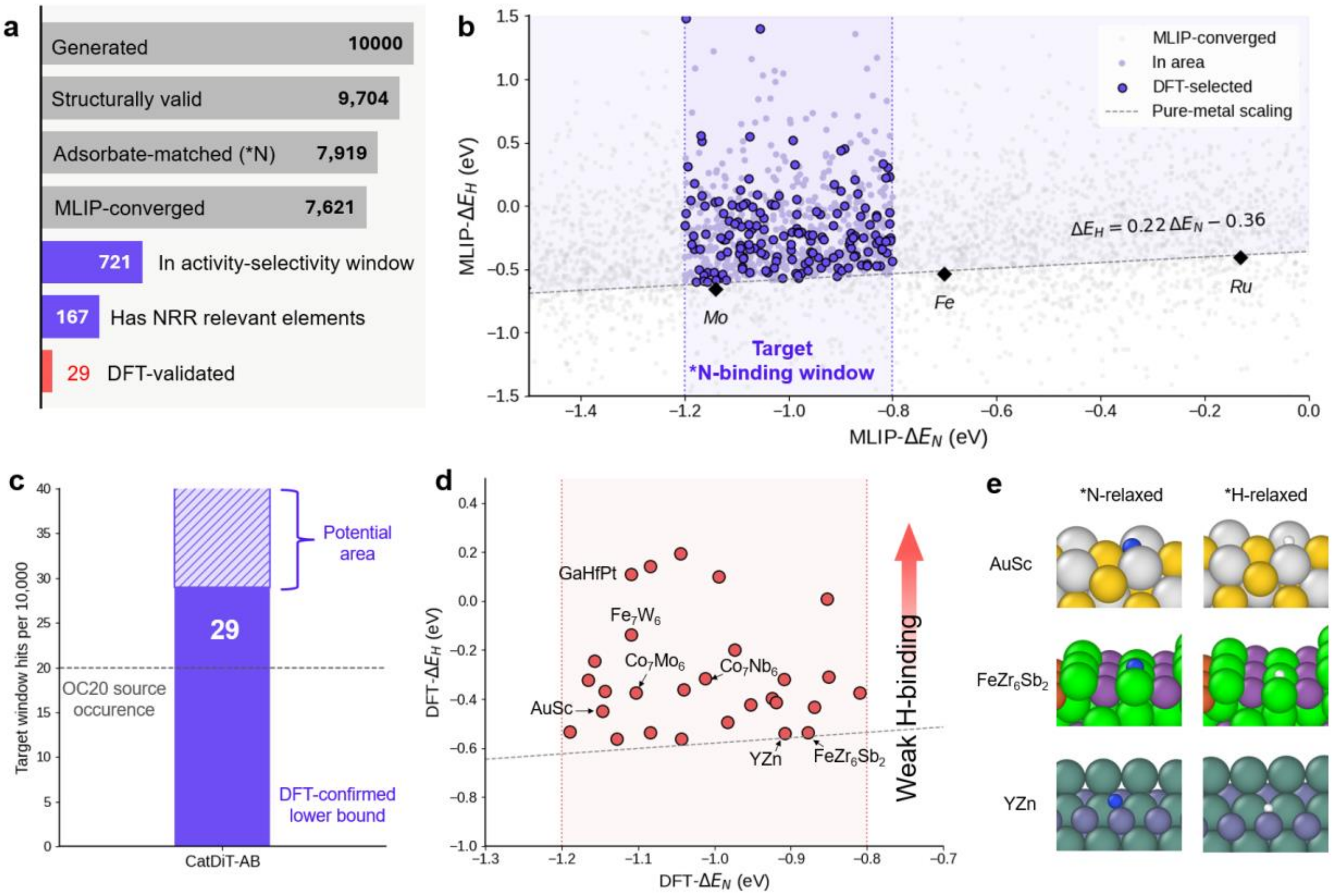

**Figure 5. Target enrichment and DFT validation of NRR catalyst candidates. a** Screening workflow for generated NRR candidates. **b** MLIP-based selection in the *N–*H descriptor space relative to the pure-metal scaling line. **c** Comparison of target-window candidates with the source dataset. **d** DFT validation in the descriptor space, where the dashed line denotes the pure-metal *N–*H scaling relation. **e** Representative *N and *H adsorption configurations of non-traditional NRR candidates.

## Discussion

In this study, we developed CatDiT as a generative framework for inverse heterogeneous catalyst design under user-defined property conditions. By learning from chemically diverse catalyst systems, CatDiT extends generation beyond narrow material classes and enables the exploration of novel candidate structures. Its compact latent-space representation of large, variable-size slab–adsorbate systems reduces the burden of atom-by-atom generation while preserving structural diversity. Combined with flow-matching-based sampling, this design enables rapid generation of large candidate pools before downstream relaxation and screening. CatDiT provides a more targeted starting point for catalyst discovery by enriching the candidate pool toward reaction-relevant regions, rather than relying on unguided generation or exhaustive enumeration. This capability is particularly valuable in heterogeneous catalysis, where promising candidates occupy only a sparse region of a vast chemical space and must simultaneously satisfy activity, selectivity, and structural constraints. Although property control remains imperfect, the framework offers a scalable route for proposing diverse, structurally plausible candidates for downstream validation. Overall, CatDiT represents a practical step toward property-directed catalyst inverse design and a useful upstream tool for accelerating reaction-specific catalyst discovery.

Nevertheless, several limitations remain. The datasets used in this study were originally constructed for energy prediction rather than generative modeling, making label imbalance and data feasibility critical factors affecting generation quality. These limitations are particularly evident in conditional generation. When a target property is underrepresented, the model struggles to generate structurally feasible outputs, as reflected by the reduced energy precision at distribution tails (Figure 4) and the correlation between training frequency and adsorbate fidelity (Figure 3). Oxide systems present additional challenges. OC22 reports only total DFT energies rather than adsorption energies, preventing adsorption-energy-based conditioning. Furthermore, the initial structures in OC20 are constructed by heuristic adsorbate placement on randomly selected valid sites rather than thermodynamically favored configurations, causing the model to learn dataset-specific placement patterns that may not correspond to catalytically relevant geometries. These limitations are present in all existing slab–adsorbate generative models built on the same datasets.

Although CatDiT-C successfully extends generation to oxide surfaces, this catalyst class presents distinct challenges. Lattice and adsorbate oxygen share the same atomic number but serve different chemical roles, and atomic tags alone do not encode bonding or coordination environments. The model also struggles to faithfully reproduce the structural regularity of oxygen-rich surfaces. These representation challenges are compounded by the prevalence of surface reconstruction, adsorbate dissociation, and desorption during DFT relaxation, which introduce noise into the learned structure–energy relationships. Collectively, these factors make oxide-based reactions, such as the oxygen evolution reaction, one of the most challenging frontiers in catalyst inverse design.

Finally, synthesizability remains a fundamental challenge in generative catalyst design, as computationally proposed structures must ultimately correspond to experimentally realizable materials. Although some candidates identified in this work match known intermetallic compositions, their structures are novel and distinct from the training data, and their thermodynamic stability and synthetic feasibility remain unverified. Addressing this challenge will require incorporating thermodynamic stability descriptors into the generation process, leveraging multi-fidelity learning with computational and experimental data, and developing quantitative synthesizability metrics.

## Methods

### Input representation

Unlike bulk crystals, catalyst structures are often difficult to describe fully solely by lattice parameters, atomic species, and fractional coordinates, commonly denoted as (L, A, F), because these features do not distinguish adsorbate atoms from slab atoms. We therefore introduce an additional atomic tag to specify the role of each atom. The OC20 and OC22 IS2RE datasets from the Open Catalyst Project, which contain initial adsorbate–slab configurations before DFT relaxation, are used in this study. Each catalyst structure with $N$ atoms is represented by the tuple $(C)$:

$$C = (L, A, F, T) = (l_1, l_2, l_3, \theta_1, \theta_2, \theta_3, a_1, f_{x_1}, f_{y_1}, f_{z_1}, t_1 \dots, a_N, f_{x_N}, f_{y_N}, f_{z_N}, t_N)$$

where $(l_1, l_2, l_3)$ and $(\theta_1, \theta_2, \theta_3)$ denote the unit-cell lengths and angles. Each atom $i$ is characterized by its atomic number $a_i \in \{1, \dots, 104\}$, fractional coordinates $f_i = (f_{x_i}, f_{y_i}, f_{z_i}) \in [0, 1)^3$, and an atomic tag $t_i \in \{0, 1, 2\}$ indicating fixed subsurface, free surface, or adsorbate atoms, respectively. The tag field is essential for catalyst structures because it distinguishes adsorbate and substrate atoms and specifies which atoms remain fixed during relaxation. For training stability, lattice lengths are normalized as $\underline{l}_k = l_k / N^{\frac{1}{3}}$, and lattice angles are expressed in radians.

In addition to the structural tuple, conditioning labels, including adsorbate identity, binding energy, and catalyst class, are extracted from the dataset metadata. Systems exhibiting adsorbate dissociation or desorption during relaxation are excluded to preserve reliable structure–energy relationships. Together with the structural tuple $(C)$, these labels constitute the complete input representation for conditional generation.

## Data preprocessing

The datasets used in this study were not designed for generative modeling. Consequently, they contain anomalous structures, such as dissociated or desorbed adsorbates, that compromise structure–energy relationships and lack sufficient structural and compositional diversity in their original data splits. We therefore applied two preprocessing steps: 1) anomaly filtering and 2) split reorganization.

For OC20, only systems with 'anomaly=0' in the raw metadata were retained. To maximize structural diversity, Val-OOD-Ads and Val-OOD-Cat were included in the training set, while Val-ID and Val-OOD-Both were used for validation and testing, respectively. The final training set contains ~383,000 clean structures.

For OC22, which includes more challenging systems, such as multi-adsorbate configurations, desorption, dissociation, and oxygen-induced surface reconstruction, only single-adsorbate structures ('nads=1' in the metadata) were retained to avoid tag ambiguity and ensure consistent adsorbate identification. This filtering reduced the dataset from 45,000 to 18,000 structures. Detailed preprocessing statistics and exploratory data analysis are provided in the Supplementary Information.

## Training, conditioning, and sampling

The training objectives follow Joshi et al.[22]. The reconstruction loss consists of cross-entropy for atom types and root mean square error (RMSE) for lattice parameters and fractional coordinates:

$$L_{rec} = \lambda_l L_l + \lambda_a L_a + \lambda_A L_A + \lambda_F L_F + \lambda_T L_T$$

We additionally introduce a tag reconstruction loss, $L_T$ (cross-entropy), to model the catalyst-specific (L, A, F, T) representation. The DiT is trained using the flow matching objective[43], in which latent vectors are linearly interpolated between noise and data. During inference, latent samples are iteratively denoised using Euler integration.

Conditional generation is implemented using CFG[19]. During training, each conditioning label is independently dropped with a probability of 50%, enabling the model to learn both conditional and unconditional generation. Discrete labels, namely adsorbate identity and catalyst class, are represented by learnable embedding tables with a null token at index 0, whereas binding energy is encoded using sinusoidal embeddings with a learnable null token. During inference, the guided prediction is computed as:

$$\tilde{v} = v_\theta(z_t, \emptyset) + w\,[v_\theta(z_t, c) - v_\theta(z_t, \emptyset)]$$

where $v_\theta$ denotes the learned velocity field, $z_t$ represents the noisy latent at timestep $t$, $c$ and $\emptyset$ signify the condition and null embedding, respectively, and $w$ indicates the guidance scale.

**Evaluation metrics**

Although many generative models for crystalline materials are evaluated using the stable, unique, and novel (SUN) framework, catalyst structures require additional considerations. Our evaluation therefore addresses four questions: (1) Do the generated structures satisfy minimum physical criteria? (2) Are they diverse and non-redundant? (3) Do they faithfully reproduce the training distribution? (4) To what extent do they explore beyond it?

For the first question, structure validity requires all interatomic distances to exceed 0.5 Å and the cell volume to be greater than 0.1 $Å^3$. Proximity to a local energy minimum is assessed by relaxing generated structures using L-BFGS with UMA[44] (f_max = 0.05 eV/Å, maximum 500 steps); SevenNet-Omni[45] is used for oxide systems in models jointly trained with OC22. We report three relaxation metrics adapted from CatFlow: energy change upon relaxation ($\Delta E_{sys}$), convergence steps, and convergence rate.

For the second question, uniqueness is evaluated using `StructureMatcher` in `pymatgen` (stol = 0.3, angle_tol = 5°, and ltol = 0.2) after removing adsorbate atoms ('tag = 2') to isolate slab-level identity. Compositional diversity is quantified as the average pairwise Euclidean distance between Magpie compositional fingerprints computed with the `matminer ElementProperty` featurizer, reflecting the breadth of compositional space sampled by the model.

For the third question, coverage and property distributions are evaluated following the CatGPT protocol. Fingerprints are constructed by concatenating CrystalNN structural descriptors and Magpie fingerprints computed on complete structures, including adsorbates. We randomly sample 10,000 structures from IS2RE Val-ID as the ground-truth (GT) set. A GT structure is considered covered if at least one generated structure satisfies both the structural-distance threshold of 0.4 and the compositional-distance threshold of 10.0. Coverage recall is the fraction of covered GT structures, whereas coverage precision is computed in the reverse direction. Property distributions are compared using the Earth Mover's Distance (EMD; Wasserstein distance) for lattice volume, density, adsorption energy, and number of elements.

For the fourth question, novelty is evaluated using continuous distance measures rather than `the discrete classification provided by StructureMatcher`, which tends to overestimate novelty in catalyst systems. For each generated structure, nearest-neighbor distances to the training set (~383,000 structures) are computed on slab-only representations with adsorbate atoms removed. Structural novelty is quantified by $d_{AMD}$, the $L\infty$ distance between AMD vectors (k = 100), whereas compositional novelty is measured by $d_{magpie}$, the L2 distance between Magpie fingerprints.

**Computational details**

Models were trained on NVIDIA H200 GPUs. For the base CatDiT model, the VAE stage required ~460 GPU-hours (100 epochs on four GPUs), and the DiT stage required ~520 GPU-hours (200 epochs on two GPUs). Full training configurations and computational costs for all variants are provided in Supplementary Table 3. During inference, generating 10,000 structures requires ~0.16 seconds per structure, whereas batches of 1,000 structures require ~0.55 seconds per structure, excluding post hoc relaxation.

First-principles calculations were performed using the Vienna Ab initio Simulation Package (VASP 5.4.4)[46,47]. Core–valence interactions and exchange-correlation effects were described by the projector augmented-wave method[48] and the revised Perdew–Burke–Ernzerhof functional (RPBE)[49], respectively. Structural relaxation was carried out until the electronic energy difference was below $10^{-4}$ eV and the residual force on each atom was less than 0.05 eV/Å. A plane-wave basis set was truncated at a kinetic-energy cutoff of 350 eV. Brillouin-zone sampling was performed using Monkhorst–Pack meshes[50], with the in-plane k-point meshes configured as ($k_1 \times k_2 \times 1$) to ensure 25 Å < $a_n \times k_n$ (n = 1, 2) < 30 Å, where $a_1$ and $a_2$ represent the unit vector sizes in the x and y directions, respectively.

## Data availability

The raw OC20 and OC22 datasets are publicly available at https://fair-chem.github.io/. The accompanying codebase provides preprocessing scripts for the raw data.

## Code availability

The code developed in this work, including structure generation demos, is available at https://github.com/doouv/CatDiT.git.

## Acknowledgements

This research was supported by the Nano & Material Technology Development Program through the National Research Foundation of Korea (NRF), funded by the Ministry of Science and ICT (RS-2024-00450102; 50% contribution), the National Research Foundation of Korea (NRF) grant funded by the Korea government (MSIT and MOE) (RS-2025-16063688), and the Korea Basic Science Institute (National Research Facilities and Equipment Center) grant funded by the Ministry of Science and ICT (RS-2024-00404602).

## Competing interests

The authors declare no competing interests.

## Author contributions

H.D. and D.H.M. contributed equally to this work. H.D. conceived the framework, conducted computational experiments, and led manuscript preparation. D.H.M. performed the DFT calculations, associated analyses, and contributed to manuscript preparation. S.B. and J.N. secured funding and supervised the project. All authors contributed to the interpretation of the results, participated in discussions, and reviewed the manuscript.

# Supplementary Information for

# Catalyst diffusion transformer: Generative inverse design of heterogeneous catalysts

*Doo et al.*

**Supplementary Methods**

**Definition and interpretation of the near-composition threshold**

The composition distance between a generated slab and a Materials Project bulk entry is defined as the maximum absolute difference between their normalized elemental fractions:

$$d_{comp} = \max_i |x_i^{slab} - x_i^{bulk}|$$

To determine the threshold, we compared the compositions of OC20 parent bulk structures and their adsorbate-removed slabs using the same procedure. The elemental ratios were preserved in all cases, yielding a bulk-slab composition distance of zero. Therefore, the large compositional deviations observed in generated slabs are unlikely to arise solely from slab cleavage. We therefore adopted a conservative near-composition threshold that permits only small deviations from exact stoichiometry.

A composition distance of 0.03 indicates that the atomic fraction of each constituent element differs from that of the matched bulk by at most 3 percentage points. Given that the generated slabs contain an average of 71.4 atoms (median: 64), this threshold corresponds to ~2.14 and 1.92 atom-equivalents, respectively, when expressed as $N_{slab} \times d_{comp}$. Among structures near the threshold, the median atom-equivalent deviation was ~2 atoms, and 90% were within ~4 atom-equivalents. Therefore, a threshold of 0.03 permits only small compositional deviations, typically on the order of a few atoms. However, the atom-equivalent deviation is provided as an intuitive measure of compositional difference and does not represent the exact number of defects or surface atoms removed.

**Supplementary Table 1. Overview of CatDiT model variants.** Training data indicate the material scope (OC20 for intermetallic alloys and OC22 for oxide surfaces), and conditioning specifies the property labels used for CFG. Asterisks denote unconditional base models whose pretrained weights are used for conditional fine-tuning.

| odel | Training data | Conditioning |
|---|---|---|
| CatDiT | OC20 | Unconditional |
| CatDiT-plus* | OC20 + OC22 | Unconditional |
| CatDiT-A | OC20 | Adsorbate type |
| CatDiT-B | OC20 | Binding energy |
| CatDiT-C | OC20 + OC22 | Catalyst class |
| CatDiT-AB | OC20 | Adsorbate type, Binding energy |

**Supplementary Table 2. Sensitivity of $d_{amd}$ to catalyst surface variations.** Pairwise distances are calculated to evaluate the resolution of the continuous novelty metrics. Structural variations include identical compositions with different crystallographic facets, surface reconstructions, and polymorphs, whereas compositional variations involve changes in stoichiometry and elemental substitutions. The results demonstrate that $d_{amd}$ and $d_{magpie}$ are sufficiently sensitive to distinguish subtle surface-level structural differences.

| Pair | $d_{magpie}$ (×1/1000) | $d_{amd}$ |
|---|---|---|
| TiPd2 (100) vs. TiPd2 (100) | 0.0000 | 0.0000 |
| TiPd2 (100) vs. TiPd2 (212) | 0.0000 | 0.4732 |
| TiPd2 (100) vs. TiPd2 (001) | 0.0000 | 0.1709 |
| TiPd2 (100) vs. TiPd2 (100) poly. | 0.0000 | 0.2007 |
| TiPd2 (100) vs. TiPd (100) | 0.0777 | 0.3256 |
| TiPd2 (100) vs. Ti2Pd (100) | 0.1445 | 0.2770 |
| TiPd2 (100) vs. TiPd3 (100) | 0.0149 | 0.5212 |
| TiPd2 (100) vs. Ti2NiPd (100) | 0.2024 | 0.2089 |
| TiPd2 (100) vs. TiGaPd2 (100) | 2.2649 | 0.5059 |
| TiPd2 (100) vs. Al2Hf2 (100) | 2.1552 | 0.7108 |
| TiPd2 (100) vs. Ag2Bi2S4 (100) | 2.6269 | 1.2364 |
| Co2Ni2Ge2 (101) vs. Co2Ni2Ge2 (212) | 0.0000 | 0.0633 |
| Co2Ni2Ge2 (102) vs. Co2Ni2Ge2 (212) | 0.0000 | 0.0729 |
| Co2Ni2Ge2 (111) vs. Co2Ni2Ge2 (211) | 0.0000 | 0.0736 |
| RbPbCl3 (111) vs. RbPbCl3 (210) | 0.0000 | 1.4295 |

**Supplementary Table 3. Computational costs of all pretrained CatDiT variants.** Hardware specifications, optimization schedules, and total computational resources are summarized for the VAE and latent DiT stages. All pretraining and fine-tuning experiments were performed on NVIDIA H200 Tensor Core GPUs. Global steps and wall times indicate the computational cost required for empirical convergence of each conditional generation task in latent catalyst flow matching.

| | VAE-S | VAE-L | CatDiT | CatDiT-A | CatDiT-B | CatDiT-C | CatDiT-AB |
|---|---|---|---|---|---|---|---|
| **GPUs** | H200 ×4 | H200 ×8 | H200 ×2 | H200 ×3 | H200 ×4 | H200 ×4 | H200 ×3 |
| **Batch size** | 20 | 16 | 16 | 20 | 20 | 14 | 20 |
| **Epoch** | 100 | 90 | 200 | 100+100 | 100+100 | 100+100 | 100+100 |
| **Global steps** | 411K | ~168K | 1,423K | 422K+320K | 422K+240K | 422K+359K | 422K+320K |
| **Wall time** | 115h (4.8d) | ~82h (3.4d) | 260h (10.8d) | ~228h (9.5d) | ~203h (8.5d) | ~219h (9.1d) | ~228h (9.5d) |
| **GPU hrs** | 460h | ~656h | 520h | 799h | 812h | 876h | 800h |

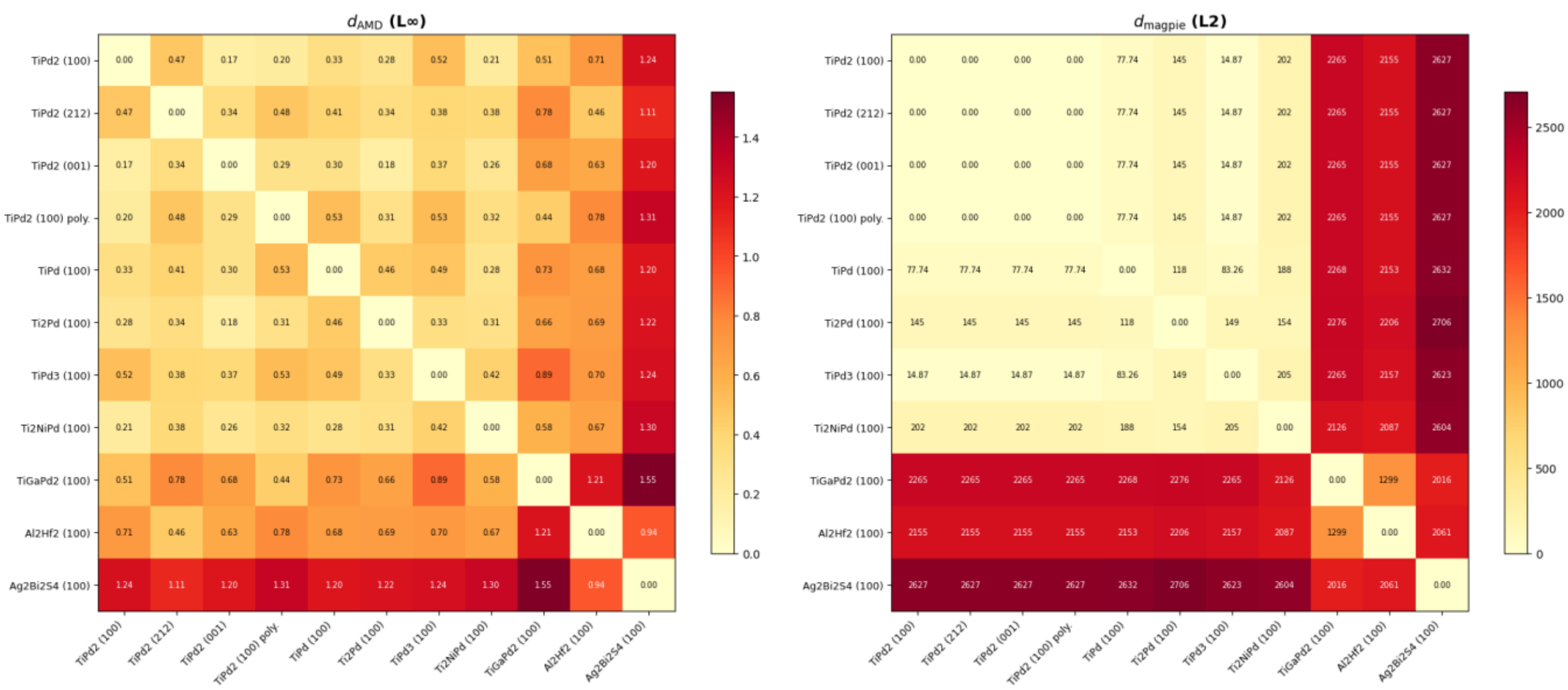

**Supplementary Figure 1. Visualization of the sensitivity of $d_{amd}$ as distance matrices.** Pairwise distance matrices are used to evaluate the resolution of the continuous novelty metrics. Left: Pairwise structural distances ($\boldsymbol{d_{AMD}}$ based on the $\boldsymbol{L_\infty}$ norm) for representative catalyst slab configurations. Right: Pairwise compositional distances ($\boldsymbol{d_{magpie}}$ based on the $\boldsymbol{L_2}$ norm) for the same structures. Color intensity reflects the magnitude of structural and compositional differences, demonstrating the ability of $\boldsymbol{d_{AMD}}$ to distinguish subtle surface-level atomic reconfigurations, crystallographic facets, and polymorphs that share identical compositions ($\boldsymbol{d_{magpie} = 0}$).

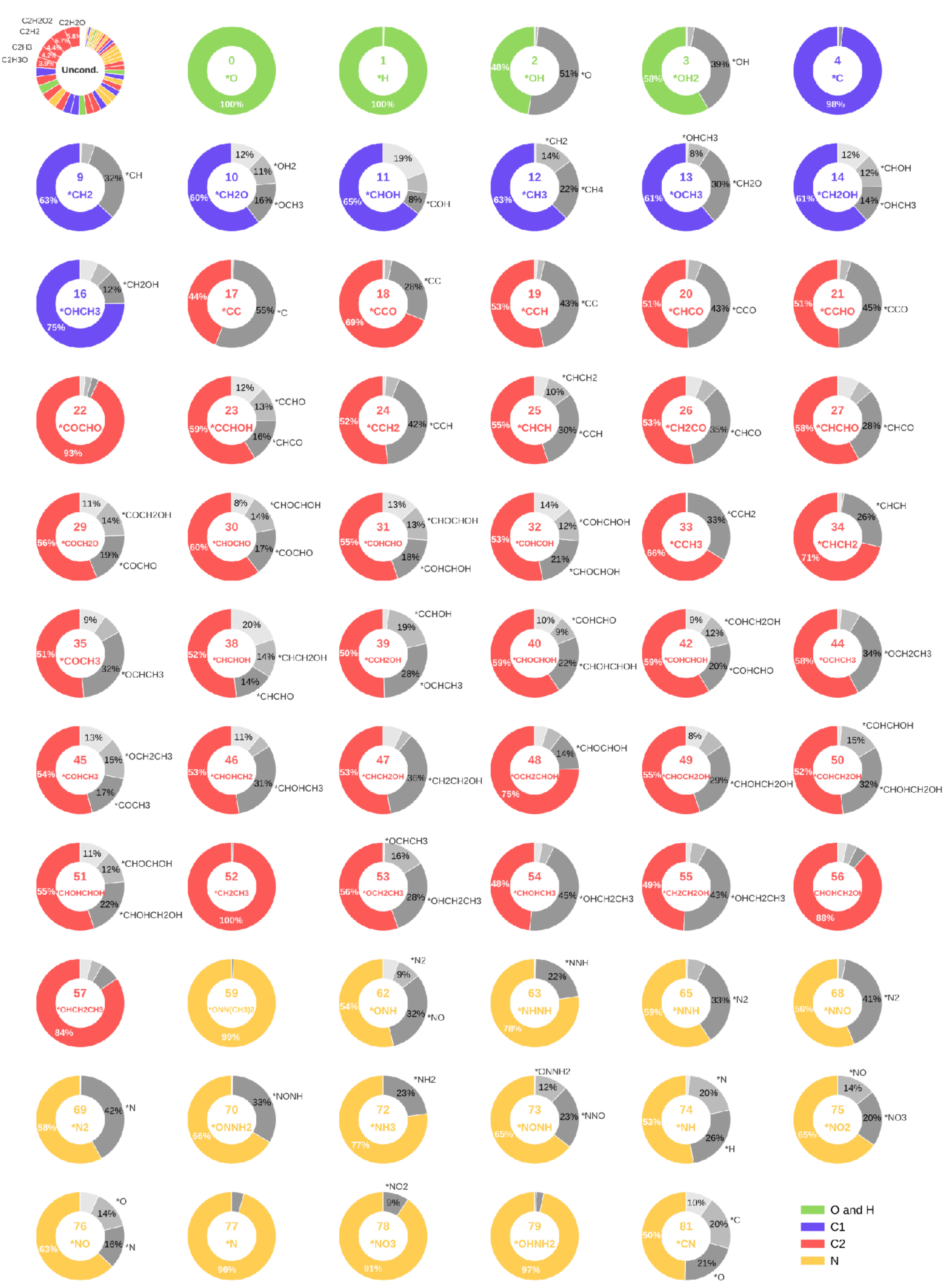

**Supplementary Figure 2. Conditional generation results for 64 adsorbates.** Pie and donut charts summarize the generation fidelity for the 64 adsorbate labels retained from the OC20 training set. Each panel shows the exact match rate for the target adsorbate and the frequencies of misassigned labels. Most misassignments occur among geometric or constitutional isomers, such as variations in $C_1/C_2$ connectivity or nitrogen-hydrogen ratios, indicating that latent space captures chemically plausible molecular topologies even when the target condition is not satisfied.

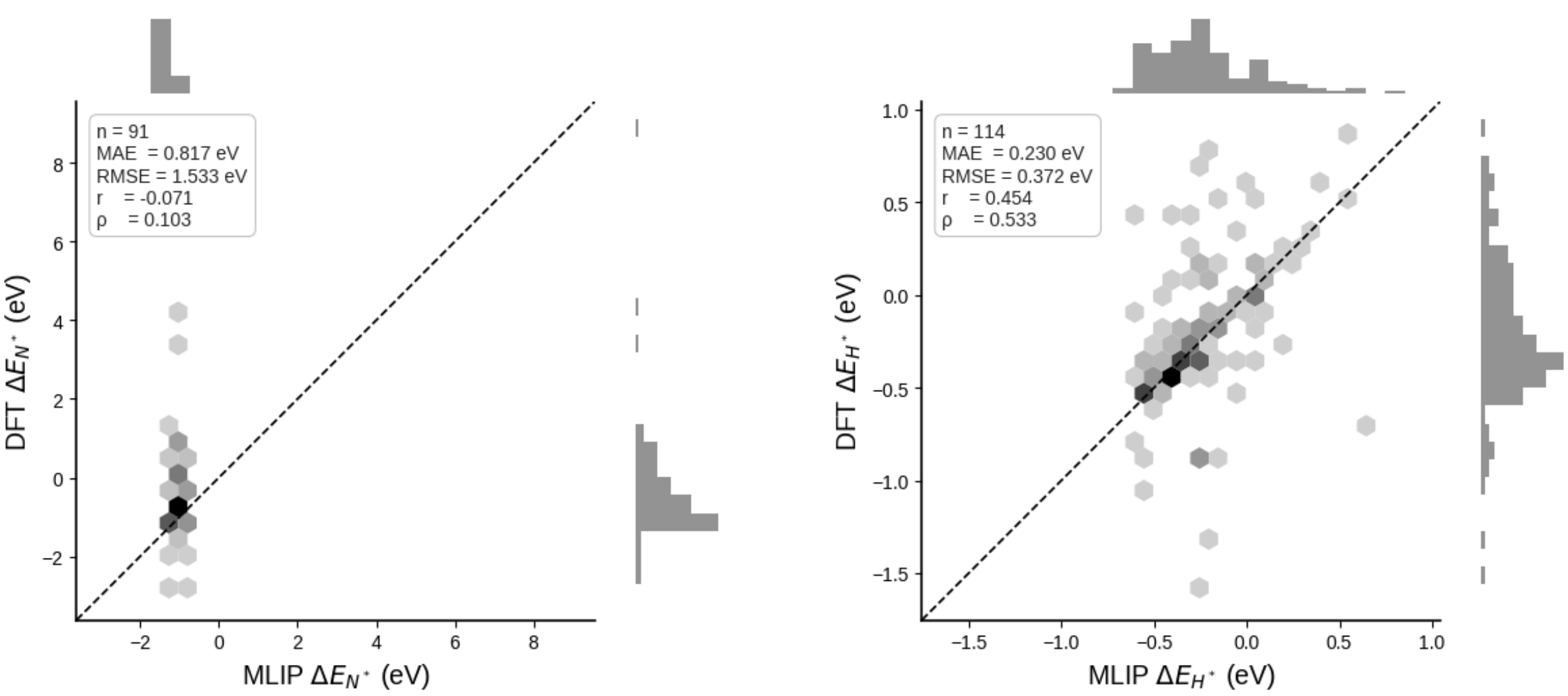

**Supplementary Figure 3. MLIP-DFT comparison for generated NRR candidates.** MLIP-predicted adsorption energies are compared with DFT reference values for adsorbed nitrogen ($\Delta E_{N^*}$, left) and hydrogen ($\Delta E_{H^*}$, right). Marginal histograms show the corresponding energetic distributions. Prediction accuracy is quantified by the mean absolute error (MAE), root-mean-square error (RMSE), and correlation coefficients ($r, \rho$). The observed deviations, particularly for $N^*$ adsorption, highlight the need for a multi-tier validation strategy in which MLIP serves as an efficient candidate prefilter before DFT validation**.**